\documentclass{article}

\usepackage{arXiv}

\usepackage[utf8]{inputenc} 
\usepackage[T1]{fontenc}    
\usepackage{hyperref}       
\usepackage{url}            
\usepackage{booktabs}       
\usepackage{amsfonts}       
\usepackage{nicefrac}       
\usepackage{microtype}      
\usepackage{lineno}
\usepackage{xcolor}
\usepackage{graphicx}
\usepackage{mathtools}
\usepackage{amssymb}
\usepackage{algpseudocode}
\usepackage{algorithm}
\usepackage{upgreek}
\usepackage[square,numbers]{natbib}
\title{Controlling Earthquake-like Instabilities using Artificial Intelligence}

\author{
  Efthymios Papachristos$^1$, Ioannis Stefanou$^{1}$ \thanks{Corresponding author; email: ioannis.stefanou@ec-nantes.fr} 
  \\
  $^1$ Institut de Recherche en G\'enie Civil et M\'ecanique, Ecole Centrale de Nantes \\ 1 Rue de la No\"e, Nantes 44321, France\\}
  
\begin{document}
\maketitle
\vspace{-30pt}
\begin{abstract}
Earthquakes are lethal and costly.
This study aims at avoiding these catastrophic events by the application of injection policies retrieved through reinforcement learning.
With the rapid growth of artificial intelligence, prediction-control problems are all the more tackled by function approximation models that learn how to control a specific task, even for systems with unmodeled/unknown dynamics and important uncertainties. Here, we show for the first time the possibility of controlling earthquake-like instabilities using state-of-the-art deep reinforcement learning techniques. The controller is trained using a reduced model of the physical system, \textit{i.e,} the spring-slider model, which embodies the main dynamics of the physical problem for a given earthquake magnitude. Its robustness to unmodeled dynamics is explored through a parametric study. Our study is a first step towards minimizing seismicity in industrial projects (geothermal energy, hydrocarbons production, CO2 sequestration) while, in a second step for inspiring techniques for natural earthquakes control and prevention.  
\end{abstract}

\section{Introduction}

The debit in human lives due to earthquakes is terrifying (one of the deadliest phenomena in nature) and so is its economic impact \cite{anbarci2005earthquake,CRED2015}.
Earthquake is a risk that has to be mitigated and so is the risk by induced/triggered seismicity from industrial projects (\textit{e.g.}, during geothermal energy and hydrocarbons production or CO2 squestration) \cite{terakawa2012high,ellsworth2019triggering,bhattacharya2019fluid,schultz2020hydraulic}.
Advances in artificial intelligence (A.I.) over the last years are spectacular. 
The question posed here therefore is: 
Could A.I. help in mitigating earthquakes?

Horrifying as this catastrophe may be, its short to medium time-scale prediction is still debatable in the scientific community.
A recent study followed a different approach. 
It built on the idea that ``the best prediction of a system's behavior is the control of it" \cite{stefanou2019controlling}. 
The author prove mathematically that the fault system is controllable and observable and that earthquakes could be avoided by injecting fluid inside the critically strained fault under specific protocols. 
The arguments behind this study are backed by field studies such as the pioneering work by \cite{raleigh1976experiment}, who conducted field-experiments to control seimicity and recent studies by \cite[among others]{guglielmi2015seismicity,cappa2019stabilization,kwiatek2019controlling, tzortzopoulos2021absorbent} that prove through systematic well-monitoring that a fault can also slip in a stable manner, that is not accompanied by abrupt seismic energy release (aseismic slip).
This new approach, converted earthquake prevention into a problem of control theory \cite{vardulakis1991linear, khalil2002nonlinear}. 

However, considering the non-continuous observations one might obtain from the monitored fault, the problem gets more complex and drifts towards the domain of discrete control \citation{discretecontrol}. 
Over the last years, artificial intelligence (A.I.) has met massive 
development and offers an alternative to discrete controllers. Its branch that excels in such control problems is Reinforcement Learning \cite{sutton1992reinforcement,lewis2012reinforcement,kiumarsi2018Optimal}.

%

Although the domain in geophysics and fault mechanics is well developed and rich, there is a huge gap between this knowledge and the knowledge acquired in the domains of mathematics and control of unstable systems. This paper is a central step for bridging this gap. 
We build, for the first time, an earthquake prevention controller based on A.I. .
More specifically, the design assumes a controlled pump that injects or withdraws fluid into a critically strained fault in order to drive it to the next stable point on a constant velocity (tracking), orders of magnitude smaller than the seismic one. 
We set up a reduced order model \cite{obinata2012model}. 
For this purpose we embody the slider analogue which reduces the dynamics of earthquake-like instabilities. 
Unmodeled dynamics and uncertainties (\textit{e.g.}, high frequency dynamics, multiphysics couplings, friction mechanisms, heterogeneities etc.) are considered as perturbations, which, the design A.I. controller stabilizes (robustness). 
The controller is trained through state-of-the-art Deep Reinforcement Learning (DRL) to select the needed pressure increment at each step of the procedure. 
The findings of this research 
open new perspectives on the domain of fault mechanics and earthquake risk mitigation.

\section{Fault System Environment and Model Reduction}
We assume a fault of the size of $L=5$ km at 5 km depth, at the limits of the seimogenic zone. 
Due to the far field tectonic movement (i.e., tectonic plates movement), the rock masses adjacent to the fault, tend to move apart one another with a far-field velocity $v_\infty$ which lays in the order of few centimeters per year along the fault plane. 
Due to the far-field movement, elastic deformation and, therefore, elastic strain energy build up inside the mobilized rock-mass.
The majority of this movement (slip) is accomodated in a zone of finite thickness called the Principal Slip Zone (PSZ) \cite{ben2003characterization, platt2014stability, scholz2019mechanics}. 
Due to the very small thickness of this zone compared to the fault system, it is reduced to a frictional interface.
The slip of the mobilized rock mass is partly restrained by the frictional force at the interface between the adjacent moving rock-masses.
The evolution of the apparent frictional resistance of the fault gouge depends mainly on cummulative slip, $y$, but also on other parameters such as slip-rate, $\dot{y}$, the so-called state \cite{dieterich1981constitutive}, pore fluid pressure $P_f$ and other physico-chemical processes \cite[among others]{reches2010fault,scholz2019mechanics,dieterich2015modeling,rattez2018importance,sulem2009thermal,veveakis2013failure,brantut2012strain}.
When a friction drop takes place and the resulting resistance force drop per slip increment is larger in absolute terms than the system's stiffness, a rapid, unstable, seismic slip occurs \cite{dieterich1978time, kanamori2004physics,scholz2019mechanics}. 
Otherwise, the slip (and thus the fault system) is slow.
These two cases characterize unstable and stable equilibria respectively in the sense of Lyapunov stability \cite{lyapunov1992general, stefanou2019controlling}.
Thus, the main feature which governs the response of the fault system is the apparent frictional behavior of the interface and its elasticity.
This justifies the use of the spring-slider model as reduced model for designing robust and efficient controllers.

The apparent frictional properties of the fault core interface can be estimated from experiments and in-situ measurements. 
We will consider typical fault properties as summarized in Table 1 of \textit{Supporting Information}. 
Note that here we assume an averaged drop in the coefficient of friction of $\Delta \mu = 0.2$ , which corresponds to a $10$ (MPa) drop in shear stress. This serves as a severe instability case to test the efficacy of the proposed method (usual stress drop values fluctuate around 3 (MPa) when averaged at the fault scale \cite{sibson2011scope}).
The fluid pressure along the fault is considered to be homogeneous and fluid diffusion is disregarded only for simplicity.

After accumulating strain for a long period of time, the fault is at the verge of instability. 
If intervention is made, the radiated energy from the unstable slip will correspond to an earthquake of magnitude 6.
Considering a reduced model for the fault system as explained in Section \textit{Fault System Model} of \textit{Supporting Information}, we will train a controller to try and control this slip aiming at imposing a constant velocity (see Section \textit{Tracking} in \textit{Supporting Information}), called hereafter the \textit{design velocity}, $v_d$. This velocity is chosen to be about two orders of magnitude smaller than the peak seismic velocity, \textit{i.e.}, $0.0028$ instead of $0.38$ (m/s). By driving the fault at design velocity, $v_d$, we ensure that the radiated energy will remain low and that the seismic energy will be dissipated smoothly on the fault zone aseismically. The whole procedure will last an operational time, $t_{op}$.

\section{A.I. Controlled Fault}

\begin{figure*}
\centering
\includegraphics[width=\textwidth]{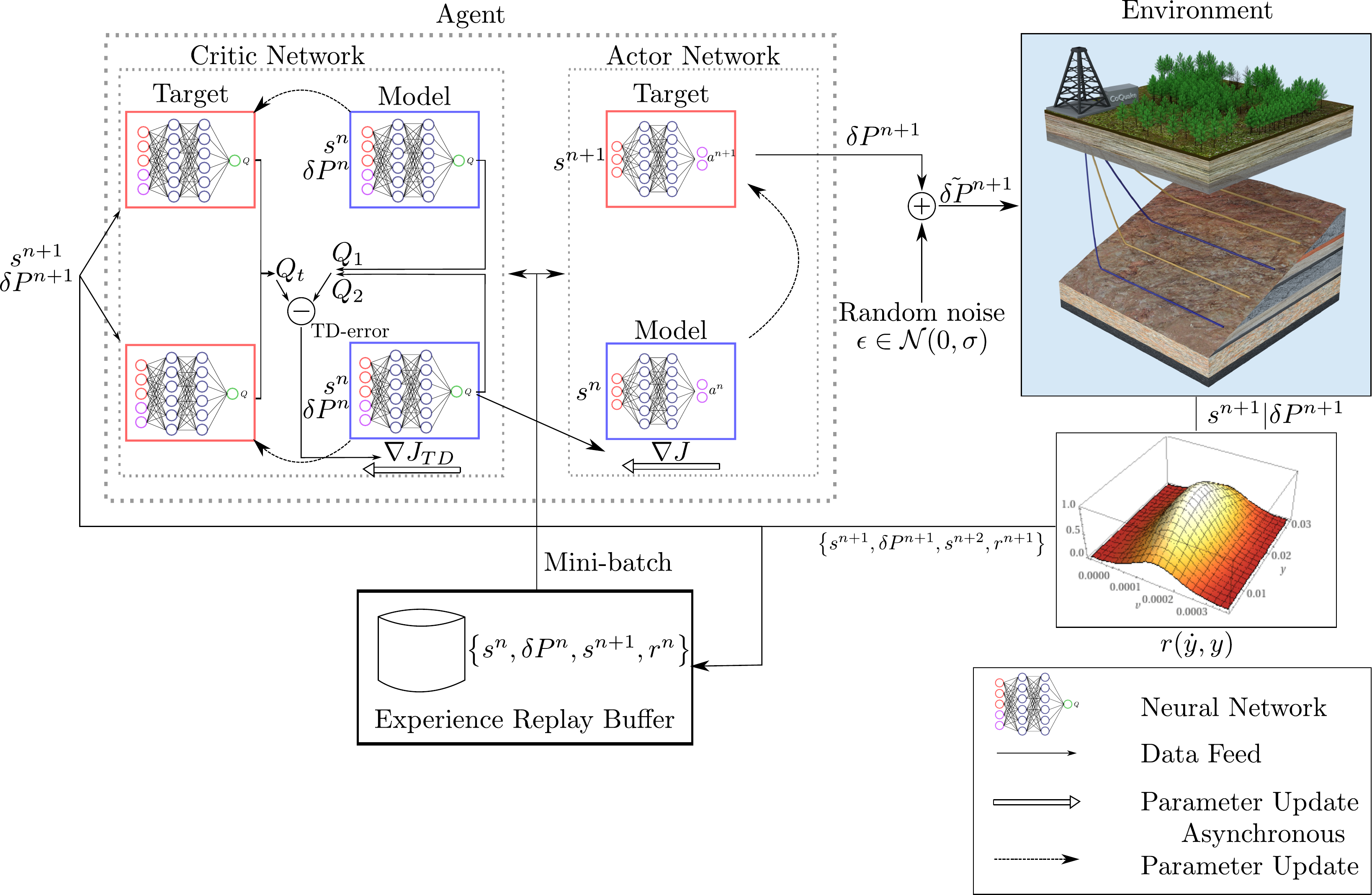}
\caption{Fault-A.I. controller diagram. The training procedure is the following. A batch of transitions consisting of current state, $s^n$, action taken $\delta P^n$, next state, $s^n+1$, reward, $r$, is fed to the TD3 model. From the next state, the Target Actor takes the next action in the environment which is evaluated by a reward function. The Target Critic Twins, using the next state along with the next action taken, compute the target Q-value, $Q_t$. The Model Critic twins calculate the Q-values at the current state $Q1,Q2$. The Temporal Difference (TD) error is used to update the Model Critics. Finally, the output of the Model Critics is used to update the Model Actor net. The system is learning in asynchronous way, the Critic-Targets and Actor Targets (in red box) are delayed and soft-updated every second step while the actor and critic targets (in red box) are updated at every step. The final optimal policy that will be used for the injection pump  after the training is the Model Actor net.}
\label{f:sketch}
\end{figure*}

In this work, the problem of controlling fault's slip will be tackled through extensive use of DRL. DRL is optimized for solving time-resolved optimal control problems (selecting the best action per step to achieve the optimal result).
The main components of the framework used are: 
\begin{itemize}
\item The \textit{environment}: The fault system described in the previous section (see \textit{Supporting Information} for detailed description of the reduced model used). 

\item The \textit{agent}: The pressure controller connected to the fluid injection pump.

\item The \textit{action}: The pressure increment $\delta P$ injected (or withdrawn) by the pump at each time-step.

\item The \textit{observations}, returned by the environment as an outcome of the latest agent's action. The observations used here are displacement $y$, velocity $\dot{y}$ and the fluid pressure $P$. These values correspond to averaged, expected values over a large region of the fault. In practice, the kinematic observations may be obtained by placing sensors (\textit{e.g.}, accelerometers) in the fault region or through satellite imaging methods \cite[among others]{avouac20062005, karimzadeh2013interseismic, walters2011interseismic, fattahi2016insar, cavalie2008measurement}. If $y,\dot{y}$ are not available by the sensors, observers can be designed \cite{khalil2002nonlinear}.
The pressure observations on the other hand, may be obtained either explicitly at the pump, or implicitly, through the injected fluid volume.

\item The \textit{reward}: A value returned by the environment at the end of each step to evaluate agent's $\delta P$ increment selection to achieve a predefined target.

\end{itemize}

The training procedure consists of letting the agent try several episodes of the scenario and gradually converging to better and better behavior. This behavior is called \textit{policy}. The policy is evaluated by a set of rewards the agent is gaining on each time-step, based on how good the action taken was. Here, the evaluation is done based on how close the system is brought to the design velocity, $v_d$, on each step. The process is repeated until the agent converges to a high-reward, optimal policy.

The DRL model selected for the controller is the Twin Delayed Deep Deterministic Policy Gradient (TD3) and it's the current state-of-the-art in reinforcement learning on continuous action spaces \cite{fujimoto2018addressing}. The model is shown in Figure~\ref{f:sketch}. More details about the model and training can be found in Section \textit{DRL Controller and Training} of \textit{Supporting Information}. Although the action and observation spaces are continuous, the environment is based on discrete sampling to simulate the feedback of a sensor in a realistic case (digital control). That is, the controller takes inputs on a discrete time interval $\Delta t$ and acts accordingly. Note that this task is more challenging than continuous feedback systems as the agent only acts in specific timestamps and has to take a precise action. In between, the pressure is kept constant (zero order hold).

\section{Avoiding Seismicity}
\label{sec:results}

\begin{figure*}[h]
\centering
\includegraphics[width=1\textwidth]{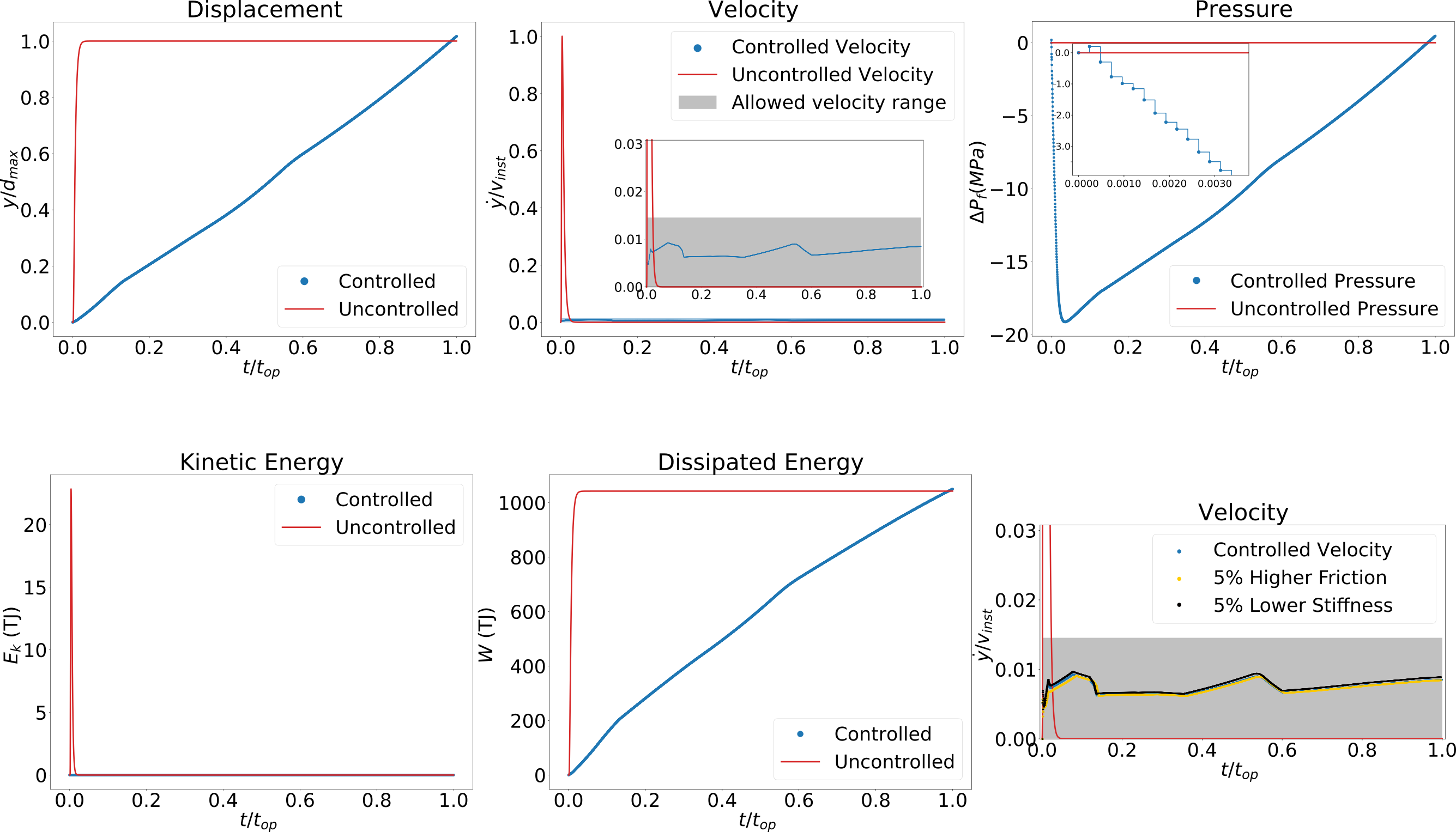}
\caption{Comparison between controlled (blue dots) and uncontrolled (red line) fault system's behavior.
(Top-plate) 
The uncontrolled case reaches the next stable point almost instantly, while in the controlled case the displacement is following a prolonged quasi-linear response. The related velocity response is following a pulse-like behavior for the uncontrolled case but a quasi-constant velocity $\dot{y}$ $\approx$ $v_d$ for the controlled case. Although there are small fluctuations on the velocity response it always remains inside the allowed velocity range (here, $\pm 1 v_d$). 
For the uncontrolled case there is no differential pressure (no fluid is injected/withdrawn), while, in the controlled case, the controller first injects water to mobilize the fault according to $v_d$ and then enables influx to avoid unwanted fast sliding.
(Bottom plate) The pulse-like behavior in velocity of the uncontrolled case, is reflected to kinetic energy responses while the dissipated energy, following displacement, is reaching the maximum value shortly. Contrary to that, in the controlled case, the kinetic energy remains low and close to 0.5 (TJ) while the dissipated energy increases quasi-linearly reaching the same final value as the in uncontrolled case.
For different values of rock stiffness and interface friction the velocity comparison plot depicts the robustness of the controller on 5\% variation in elastic properties and friction coefficient.
}
\label{f:Results}
\end{figure*}

The response of an idealized fault system without the use of the A.I.-controlled pump is shown in Figure~\ref{f:Results} with the solid red lines. In this numerical example, the instability is caused after an aseismic slip of $y \approx$ 10 (mm) reaching a maximum slip velocity $v_{in} = 0.38$ (m/s). 
The fault slips fast until the new steady state is reached after a slip $d_{max} =1.667 $ (m). The kinetic energy during the slip follows the velocity pulse and reaches $E_k \approx 22.5$ (TJ), while the total dissipated energy is at the order of $W \approx 1000$ (TJ). 
This behavior corresponds to an earthquake of Magnitude 6 \cite[among others]{kanamori2004physics}.

If, instead, the A.I. controller is activated, the fault is slipping at an almost constant slip rate $v_d = 0.0028 $ (m/s). This is reflected into slipping kinetic energy which is kept practically constant with a maximum of $E_k \approx 0.5 $ (TJ). This value is three orders of magnitude lower than the in the uncontrolled system. 
The dissipated energy is quasi-linearly increased reaching $W \approx 1000$ (TJ).
As a result, we could say that the controlled system quasi-statically and the slip is aseismic. 
From a mathematical point of view, the closed-loop, controlled system, is now stable. 
Earthquake is avoided.

\section{Discussion}
In this work we built an A.I. controller that prevents the seismic-slip instability and the associated abrupt energy release, \textit{i.e.}, it prevents earthquakes. The fault system is stabilizable and controllable and thus a controller can be designed for any set of initial conditions. 
The A.I. controller is using discrete observations from the environment to adjust the fluid pressure inside the fault at a given finite time-step. 

The A.I. controller was based on a reduced model that can represent the main dynamics of the physical system. The main dynamics of this problem that we want to control are related to the earthquake instability. As a result, poroelastic effects were omitted and the realistic fault geometry is simplified to an idealized system perfectly oriented for slip. 
Moreover, we focused in controlling a single seismic event, not to model it's long-term repeatability. Hence a slip-weakening law was introduced.
Model reduction is common in control theory \cite{obinata2012model}, where any unmodelled dynamics and uncertainties are supposed to be dissipated by the controller.
Here, we provide a proof of concept for earthquake control using Deep Reinforcement Learning and therefore we limit the complexity of the underlying physical system.

\begin{figure*}[h]
\centering
\includegraphics[width=\linewidth]{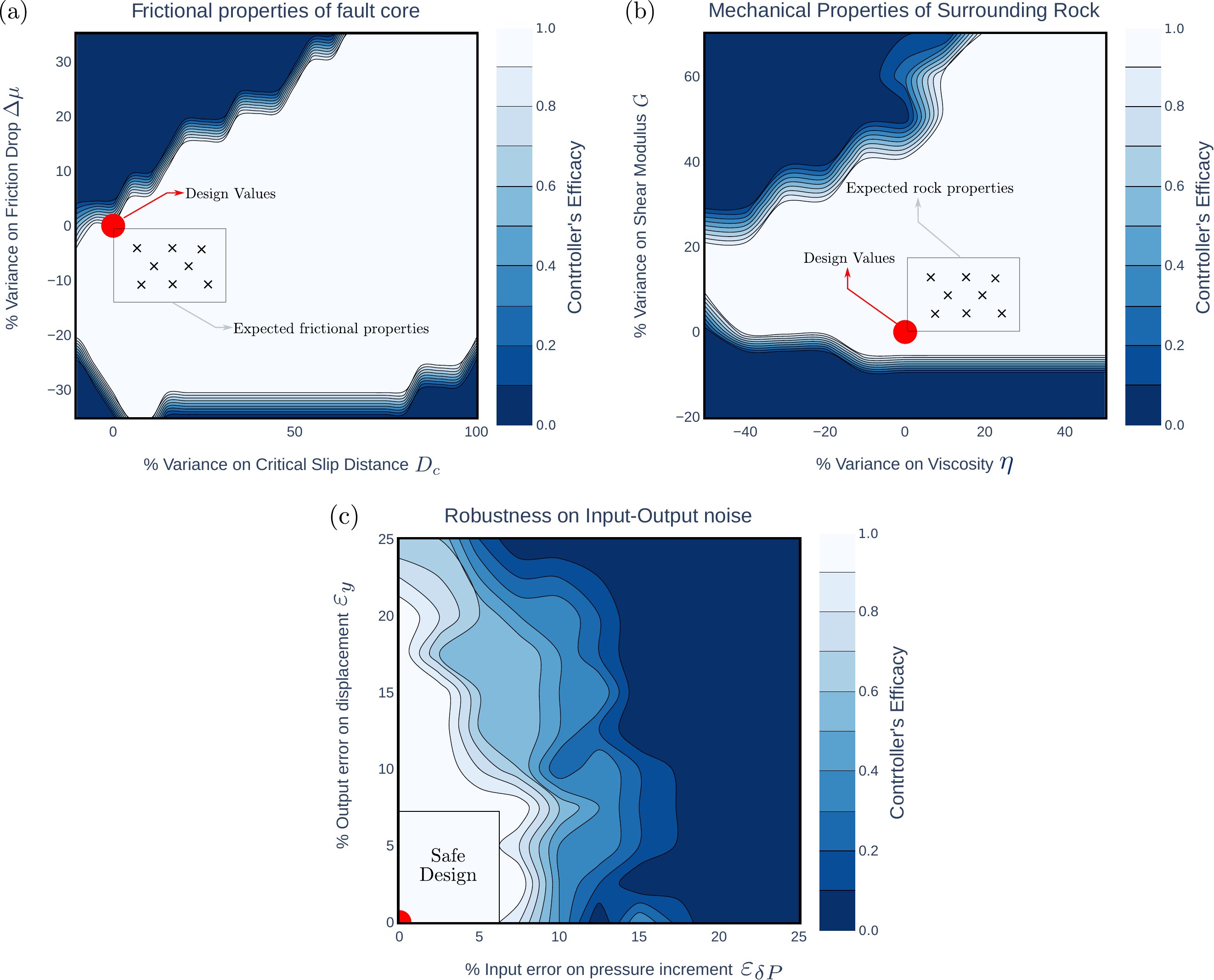}
\caption{Robustness analysis for the A.I. controller. (a) Efficacy of the controller on variations of the frictional properties of the fault core, more specifically, the critical slip distance, $D_c$, and the friction drop $\Delta \mu$. (b) Efficacy of the controller on variations of the mechanical properties of the surrounding rock, such as the shear modulus, $G$, and the apparent viscosity, $\bar{\eta}$. 
Both studies, (a) and (b), show that the exact values for the frictional and mechanical properties of the fault system are not necessary for an effective controller. If the worst-case, with respect to fault's stability, parameters-set is used for the design (red circle), the controller will remain robust for a large range of uncertainties.
(c) Efficacy of the controller on errors in displacement readings (output noise) and pressure increment application (input noise). As a first approximation, a region of admissible input-output, sensor errors for safe design can be identified and bounded at $7.5\%$ error in both input and output noise.}
\label{f:robustness}
\end{figure*}

Nevertheless, to test the applicability of the derived controller to more complicated conditions, a robustness analysis was conducted. 
Variations on mechanical properties of the surrounding rocks as well as frictional properties of the formation are expected on field applications. The analysis showed that the controller is functional for an increase up to $5\%$ of apparent friction drop in the fault core, $\Delta \mu$, or a $5\%$ decrease on of shear modulus of the surrounding rock, $G$, and a decrease up to $40\%$ on surrounding rock's viscosity, $\bar{\eta}$ , see Figures~\ref{f:robustness}a, \ref{f:robustness}b. 
What is even more interesting, is that the controller remains perfectly functional in large variances that favor the stability of the system such as the increase modulus $G$, increase of critical slip distance $D_c$ and decrease of apparent friction drop $\Delta \mu$. 
This justifies the design of such a controller, through the \textit{worst-possible scenario} perspective: 
The exact values of the frictional and mechanical properties of the fault system are not necessary for an effective controller. If the controller is designed for the least favorable conditions expected, regarding (in)stability, it could compensate relatively large variations and remain robust in driving the fault to a new stable equilibrium aseismically. 

Furthermore, to minimize unexpected risk, we considered the possibility of obtaining corrupted data errors from the sensors, in terms of displacement reading, $\mid \varepsilon_{y}\mid$ as well as errors (perturbations) at the input pressure, $\mid \varepsilon_{\delta P} \mid$. Both  $\varepsilon_{y}$ and $\varepsilon_{\delta P}$
correspond to the standard deviation of a Gaussian type noise around the original value, such that the final observed displacement is given by $\tilde{y} = y (1 \pm \varepsilon_{y})$ and the final applied pressure increment by $\tilde{\delta P} =\delta P  (1 \pm \varepsilon_{\delta P})$ respectively. Furthermore, due to the stochastic nature of the noise, a series of 5 different sets of parametric tests were used for the robustness analysis.
The results of the robustness analysis are shown in Figure~\ref{f:robustness}c. 
Considering only output errors in displacements, the controller remains robust for random input errors (displacement readings) up to $\mid \varepsilon_{y}\mid = 25\%$, while considering only output (pressure increment) errors the controller can safely handle variations up to $\mid \varepsilon_{\delta P} \mid = 7.5\%$.
For a combined input-output error, a safety margin emerged, bounded by $\mid \varepsilon_{y}\mid=7.5\%$ and $\mid \varepsilon_{\delta P}\mid=7.5\%$. 
This safety margin can serve as a first order approximation on the controller's design and application.

Overall, this work presented a novel method, based on state-of-the-art A.I. techniques, specifically DRL, to control natural, induced or triggered fault instabilities and thus to prevent earthquakes.
The method was tested in an reduced model of a fault system perfectly oriented for slip, which can give earthquakes of $M_w=6$. 
Nevertheless, the A.I. controller managed to drive the fault in a slow, aseismic way to a new stable equilibrium. In the same time, the stored elastic energy was safely mitigated. Furthermore, it was shown that the A.I. controller does not require the knowledge of the exact values of frictional and mechanical properties (which can be difficult to measure \textit{in-situ}), but rather, a range of expected values. 
The method itself and the findings of this study could find fertile ground for minimizing seismic risk in industrial projects as well as for avoiding potential natural earthquakes in the future. 

\section*{Acknowledgments}
This work was supported by the European Research Council (ERC) under the European Union Horizon 2020 research and innovation program (Grant agreement no. 757848 Co-Quake, IS). This paper contains no data.

\bibliography{biblio}

\end{document}


\maketitle

\noindent\textbf{Contents of this file}
\begin{enumerate}
\item Methods
\item Figure S1
\item Table S1 
\end{enumerate}
\vfill

%
%

%

%
%


\newpage
\section*{Methods}
\section{Fault System Reduced Model}

\begin{figure}[h]
\centering
\includegraphics[width=\linewidth]{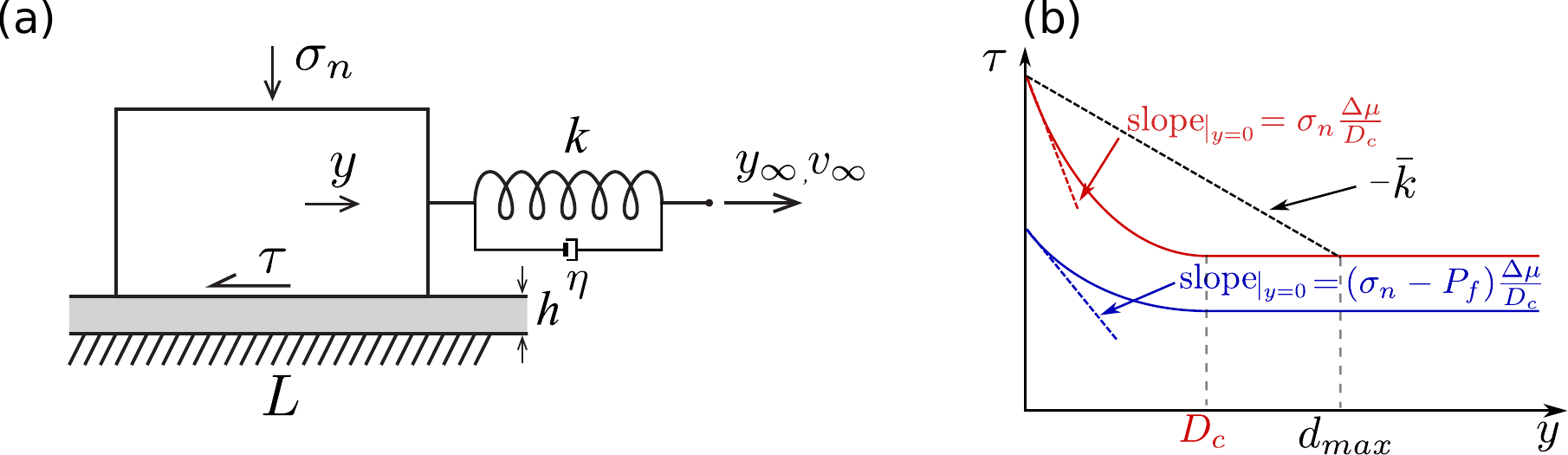}
\caption{(a) Reduced model of the fault system. (b) Coulomb frictional force evolution with slip, $y$ in presence and absence of pore pressure.}
\label{f:spring_slider}
\end{figure}

In this work, a classical, reduced order fault system analogue, which is mathematically described by a 1D spring-slider model, is considered \cite{reid1910mechanics,burridge1967model,kanamori2004physics}. 
According to this model, the mass that is mobilized during a seismic slip (the rockmass adjacent to the fault core) is represented by a rigid cubic block of size $L=5$ (km), density, $\rho = 2500$ (kg.m$^{-3})$ and mass $m \simeq \rho L^3$. 
Due to far-field tectonic movement, of constant velocity $v_{\infty}=3.18\times10^{-10}$ (m/s), 
the mobilized mass is strained, building up elastic energy. 
The stiffness of the surrounding rockmass is modeled by means of a normalized stiffness, $\bar{k}$, which is a function of the effective shear stiffness of the rockmass, $G=30\times10^3$ (MPa), and the fault's length, $L$ , as $\bar{k} = G/L$ \cite{palmer1973growth,stefanou2019controlling}. The equivalent stiffness of the elastic spring is given by $k= A\bar{k}$, with $A=L^2$ the activated fault area. 
Finally, an apparent viscosity, $\bar{\eta} = \eta/ L =10^5$ (MPa.s), of the rock mass is considered, where $\eta$ is the viscosity of a dashpot connected in a Kelvin-Voigt configuration \cite{wang2017multistable, stefanou2019controlling}. 
The reduced mechanical fault-system model is sketched in Figure~\ref{f:spring_slider}a. 

The fault core consists of a zone of localized deformation between opposite driven rock masses accomondating the majority of the slip (PSZ). This zone is most commonly characterized by a thin layer of granular gouge material, product of abrasive wear and ultra-cataclastic flow from previous slip movement \cite{scholz2019mechanics}. Its complex structure combined with the various multi-physical phenomena that take place during the pre- and co-seismic slip, lead to a variation of friction during slip. Due to its very small thickness, $h$, compared to the whole system, the fault core is mathematically studied as an interface. The apparent frictional behaviour of the fault core (\textit{i.e.}, the interface) might depend on several parameters, such as total slip $y$, slip rate $\dot{y}$, characteristic grain size evolution, diagenetic fluid expulsion, temperature, maturation and chemical phenomena\cite{reches2010fault,rattez2018importance,sulem2009thermal, brantut2012strain,veveakis2013failure}.
Considering a Coulomb friction model, the average, apparent frictional response of the interface, will depend on the coefficient of friction $\mu$, the normal stress at the given depth, $\sigma_n=100$ (MPa), and the pore pressure at the interface, $P_f=50$ (MPa), \textit{i.e.}, $\tau = (\sigma_n - P_f)\mu$. 
Several models for the constitutive behaviour of gouge materials have been proposed \cite{dieterich1981constitutive,dieterich2015modeling,aharonov2018physics}. In this work we assumed that the material's frictional behaviour is described by the following slip-weakening exponential decay model $\mu (y) = \mu_{res} \left(1 - \frac{\Delta \mu}{\mu_{res}}\right) e^{-\frac{y}{D_c}}$ \cite{andrews1976rupture, di2011fault, rattez2018importance}, 
where $\mu_{res}=0.5$ is the residual (dynamic) friction coefficient, $\Delta \mu =0.2$ is the drop in friction coefficient from static to dynamic and $D_c =10$ (mm) is a characteristic value for the gouge material describing the cumulative slip needed to transit from static to dynamic friction values. Rate-and-state laws could be used as well, but as we are interested in a single event, the slip weakening law is a reasonable assumption.

Considering a critically strained fault, the criterion for seismic slip as described in Section \textit{Fault System Analogue}, 
reads \cite{scuderi2017frictional,reid1910mechanics}:

\begin{equation}
(\sigma_n - P_f) \frac{\Delta \mu}{D_c}< - \bar{k} 
\end{equation}

\begin{table}[h]
\centering
\caption{Control parameters used in the simulations.}
\medskip
\begin{tabular}{cccc}
\hline
Parameter & Symbol & Value & Units \\
\hline
Timestep & $\Delta t$ & 0.144 & (s)  \\
Operation time &  $t_{op}$ & 600 & (s) \\
Displacement to stability & $d_{max}$ & 1.668 & (m) \\
Instability velocity & $v_{in}$ & 0.38 & (m/s) \\ 
Design velocity & $v_{d}$ & 0.0028 & (m/s) \\
\hline
\end{tabular}
\label{t:faultProp}
\end{table}

\section{Tracking}

The fault system as described in the previous section, would slip in an unstable manner, causing an earthquake. The co-seismic slip velocity, $v_{in}$, emerging from the instability reaches a maximum of $v_{in} = 0.38$ (m/s). 
The controller drives the system to the next equilibrium point, which is found at a distance $d_{max} = (\sigma_n - P_f)(-\Delta \mu) / \bar{k}$, while keeping the slip-rate at a design velocity, $v_d$, much lower than the co-seismic slip velocity, $v_{in}$ (see Table~\ref{t:faultProp}).
The slip rate of the fault is arbitrarily selected equal to a design velocity, $v_d$ and the exact position of the new equilibrium is needed to be known. The whole procedure for translating the system to a stable equilibrium will last, $t_{op}$, which is also arbitrarily selected.  
To do so, the controller adjusts the pressure inside the fault by $\delta P$ through a point injection at each time-step. The pressurization/de-pressurization increment has to achieve the appropriate resulting resistance shear stress, $\tau$, needed to satisfy a stable slip. 

Once the fault reaches a new stable equilibrium point the controller stops and stability is restored. The earthquake-like instability is avoided until the next strain energy build-up. 

\section{DRL Controller and Training}

The A.I. model used to train the controller is based on DRL. DRL lays in the field of semi-supervised learning techniques. The model is thus not trained by approximating series known data (labels). In DRL the main components are the Agent, the Environment, the Reward, the Observations and the Actions. 
The environment used here is the reduced model described in paragraph \textit{Fault system model}, and the increments in time, where actions are taken and observations are obtained, are set to $\Delta t = 0.144$ (s).
Each state is defined by the observations returned to the agent. Here, the state is defined by the cumulative slip, slip rate and the fluid pressure $P$. The agent takes actions in the environment at each step. Based on the reward function a reward is returned to the agent along with a set of observations from the environment. 
Each time the environment reaches a terminal state, it is called an end of an episode. The total number of steps per episode are $N = t_{op}/{\Delta t}$.
The agent's training is twofold, it has to be trained (a) to predict and (b) to control. More specifically it has to (a) ``learn" how to better predict state-action values Q (which will be used as labels for the policy gradient) and (b) use these Q-values to ``learn" the optimal policy. Optimal policy is the policy that leads to the highest expected return, or, cumulative reward:

\begin{equation}
J(\phi) = \mathbb{E}_{s^n\sim p_{\pi},\delta P^n\sim \pi} \left[R_o \right]
\end{equation}

\noindent where $R_o = \gamma^{n-t} r (s^n,\delta P^n)$ is the return, with $\gamma$ the discount factor, $s^n$ the state of ``time-step" $n$, $\delta P^n$ the action taken from state $s^n$ and $\pi$ the policy.

The DRL architecure used in this study is the TD3 model by \cite{fujimoto2018addressing}, as depicted in Figure 1 of the main manuscript. The original architecture was kept for the networks setup. More specifically, the Critic Target and Model nets consist of linear, feed-forward networks. The input layers hold the states and actions. They are followed by two hidden layers of 400 and 300 neurons. A ReLu activation functions is used by the first two layers. The final layer outputs the $Q(s,\delta P)$ (and $Q(s^{n+1},\delta P^{n+1})$ respectively) value. 
The Target Actor and Model actor nets follow a similar architecture, with states as input layer, followed by the same 400 - 300 neurons hidden layers and output the action, $\delta P^n$, that will be taken. The two first layers use ReLU activation functions while the last one uses a $\tanh$ activation function to account for negative and positive action values.
We used a delayed, soft update, every second step for the Target Critics and Target Actor. The soft update is done through Polyak averaging. The update rate used for the update is $\uptau = 0.005$, the discount factor, $\gamma = 0.99$ and the standard deviation for the action noise $\sigma = 0.2$.  
The model was trained in a total of $N=2\times10^6$ steps, using the following convex function of displacement and velocity as reward function:

\begin{equation}
r = e^{-1.99054 (-1 + 7142.86 \dot{y})^2 - 1.99054 (-1 + 56.8246 y)^2}
\end{equation}

\noindent This reward function is fed with step-wise velocity and displacements and gradually drives the model to it's maximum, which is the design velocity, $v_d$.

\bibliography{biblio}